\documentclass[11pt]{article}

\usepackage{amssymb}
\usepackage{mathrsfs}
\usepackage{chet}

\newcommand{\lsp}{\hspace{1pt}}
\newcommand{\llsp}{\hspace{0.5pt}}

\newcommand{\dtwo}{\ensuremath{d^{\hspace{0.5pt}2}\hspace{-0.5pt}}}

\newcommand{\dsix}{\ensuremath{d^{\hspace{1pt}6}}}

\newcommand{\st}{\mathscr{T}}
\newcommand{\sz}{\mathscr{Z}}
\newcommand{\cA}{\mathcal{A}}
\newcommand{\cB}{\mathcal{B}}
\newcommand{\cC}{\mathcal{C}}
\newcommand{\cD}{\mathcal{D}}
\newcommand{\cE}{\mathcal{E}}
\newcommand{\cF}{\mathcal{F}}
\newcommand{\cG}{\mathcal{G}}
\newcommand{\cH}{\mathcal{H}}
\newcommand{\cI}{\mathcal{I}}
\newcommand{\cO}{\mathcal{O}}
\newcommand{\vev}[1]{\langle #1\rangle}

\date{April 2016}

\title{Constraints on Perturbative RG Flows \\\vspace{5pt}
in Six Dimensions}

\author{Andreas Stergiou,$^{\!a}$ David Stone,$^{\!b}$ and
Lorenzo G.\ Vitale$^{c}$}

\affiliation{$^{a}$Department of Physics, Yale University, 217 Prospect St,
New Haven, CT 06520, USA\\
$^{b}$INFN, Sezione di Roma, Piazzale A.\ Moro 2, I-00185 Roma, Italy\\
$^{c}$Institut de Th\'eorie des Ph\`enom\'enes Physiques, EPFL, \\
Route Cantonale, CH-1015 Lausanne, Switzerland}

\abstract{When conformal field theories (CFTs) are perturbed by marginally
relevant deformations, renormalization group (RG) flows ensue that can be
studied with perturbative methods, at least as long as they remain close to
the original CFT. In this work we study such RG flows in the vicinity of
six-dimensional unitary CFTs.  Neglecting effects of scalar operators of
dimension two and four, we use Weyl consistency conditions to prove the
$a$-theorem in perturbation theory, and establish that scale implies
conformal invariance. We identify a quantity that monotonically decreases
in the flow to the infrared due to unitarity, showing that it does not
agree with the one studied recently in the literature on the
six-dimensional $\phi^3$ theory.}

\begin{document}
\maketitle
\toc

\section{Introduction}

The evolution of physical quantities with energy in quantum and statistical
field theories is described by the renormalization group (RG).  According
to the Wilsonian picture, the RG flow from the ultraviolet (UV) to the
infrared (IR) corresponds to a coarse-graining of degrees of freedom, and
should therefore be irreversible.  It is interesting to ask whether there
is any physical observable in quantum field theory (QFT) that can be
understood as the ``number of degrees of freedom'', and which decreases
along the RG flow.

This intuition has been beautifully borne out in $d=2$ spacetime dimensions
by Zamolodchikov~\cite{Zamolodchikov:1986gt}, who established that a
certain combination of two-point functions of the stress-energy tensor,
called $C$, monotonically decreases in the flow to the IR in unitary QFTs.
This so called $C$-function is stationary at fixed points of the RG, where
conformal field theories (CFTs) live, and is equal to the central charge of
the corresponding CFT there.

Soon after Zamolodchikov's work, Cardy attempted a generalization to
$d=4$~\cite{Cardy:1988cwa}, where he suggested that $a$, the coefficient of
the Euler term in the four-dimensional trace anomaly, plays the role of the
monotonically decreasing quantity. Although a general proof of the
monotonicity of $a$, commonly referred to as the $a$-theorem, was not
obtained in~\cite{Cardy:1988cwa}, significant differences with the $d=2$
case were elucidated, and further support was given to the intuition that
results similar to Zamolodchikov's should hold in any even spacetime
dimension.

Osborn~\cite{Osborn:1991gm} later analyzed the case of a unitary CFTs in
$d=4$ deformed by a set of marginally relevant operators.  By studying the
Wess--Zumino consistency conditions for the anomalous Ward indentity of
Weyl rescalings, within the formalism of the local Callan--Symankzik (CS)
equation, Osborn derived a perturbative proof of the $a$-theorem, using
also results of~\cite{Jack:1990eb}.\footnote{Although the arguments
in~\cite{Jack:1990eb, Osborn:1991gm} rely on perturbative computations
around the free theory, they can be generalized to the case where the RG
flow lies perturbatively close to any interacting CFT, weakly or strongly
coupled~\cite{Baume:2014rla, Jack:2013sha, Osborn:2015rna}.  In fact, the
CFT need not even have a Lagrangian
description.}$^{\hspace{-0.5pt},}$\footnote{A recent review of this
approach to the $a$-theorem can be found in~\cite{Shore:2016xor}.} More
specifically, an equation of the form
\begin{equation}
\label{eq:gradientflow}
\frac{\partial}{\partial \lambda^I} \hat{a} = (\chi_{IJ} + \xi_{IJ})
\beta^J
\end{equation}
was derived in~\cite{Jack:1990eb, Osborn:1991gm}, where $\hat{a}$ is a
local function of the coupling constants $\lambda^I$ of the theory, which
reduces to $a$ at fixed points, $\chi_{I J}$ and $\xi_{I J}$ are symmetric
and antisymmetric tensors respectively, defined also in terms of
$\lambda^I$, and $\beta^I$ is the beta function associated with
$\lambda^I$. By unitarity, $\chi_{I J}$ is positive-definite at leading
order in perturbation theory, as it can be related to the two-point
function $\langle\cO_I(x)\lsp\cO_J(0)\rangle$ of marginal operators. Upon
contracting (\ref{eq:gradientflow}) with $\beta^I$, one gets
\begin{equation}
\label{eq:atildeflow}
\mu \frac{d}{d \mu} \hat{a} =\chi_{I J} \beta^I \beta^J  \geq 0\,,
\end{equation}
thereby establishing the monotonicity of $\hat{a}$ along perturbative RG
flows. The inequality is saturated only if $\beta^I=0$.

Recently, Komargodski and Schwimmer \cite{Komargodski:2011vj} demonstrated
that $a$-theorem holds true beyond perturbation theory in $d=4$, more
specifically that $a_{\text{UV}} - a_{\text{IR}}$ must be positive in
unitary theories.  Their argument relies on dispersion relations for
four-point scattering amplitudes for the dilaton, i.e.\ the background
source for the trace of the stress-energy tensor.  The connection between
the non-perturbative and perturbative arguments was made
in~\cite{Baume:2014rla}, where it was shown how equation
(\ref{eq:gradientflow}) can be extended beyond leading order by employing
the dilaton effective action.

A question closely related to the $a$-theorem is that of the relation
between scale and conformal invariance, in particular whether scale
invariant field theories (SFTs) enjoy the full conformal symmetry under the
assumptions of locality and unitarity.  Polchinski proved the equivalence
$\text{SFT}=\text{CFT}$ in $d=2$~\cite{Polchinski:1987dy}. In $d=4$,
perturbative checks were performed in~\cite{Polchinski:1987dy} as well
as~\cite{Dorigoni:2009ra}, and general perturbative arguments were later
given in~\cite{Luty:2012ww, Fortin:2012hn}.  Beyond perturbation theory in
$d=4$ conditions for the equivalence of scale and conformal invariance have
been analysed in~\cite{Dymarsky:2013pqa}.

Due to the importance of the $a$-theorem and its consequences for the
structure of QFTs, it is of great interest to continue the exploration of
these ideas to higher spacetime dimensions, in particular $d=6$.  Some
important results have been obtained in~\cite{Elvang:2012st,
Cordova:2015fha}, but in this work we will focus on the approach pioneered
by Osborn in~\cite{Osborn:1991gm}, which relies on the local CS equation.
This formalism was recently generalized to $d=6$~\cite{Grinstein:2013cka},
where complications arise due to the large number of terms that have to be
considered in the Weyl anomaly.

In the present work we study the RG flow in the proximity of a
six-dimensional CFT by deforming it by a set of marginally relevant
operators $\cO_I$,
\begin{equation}
\label{eq:perturbedtheory}
S[\Phi,  \lambda] = S_{\rm CFT}[\Phi] + \int \dsix x\, \lambda^I
\mathcal{O}_I(x)\,.
\end{equation}
For simplicity, we assume that relevant operators of dimension two and four
are absent from the theory. We plan to include their contributions in
future work.

By analyzing the Wess--Zumino consistency conditions in the context of the
local CS equation, we will be able to identify a function of the coupling
constants, $\hat{a}$, satisfying an equation analogous to
(\ref{eq:gradientflow}), thereby proving the $a$-theorem in perturbation
theory. In fact, we find a one-parameter family of functions,
$\hat{a} + \lambda \hat{b}$, satisfying an equation of the form
\begin{equation}
\mu \frac{d}{ d \mu} \big( \hat{a} + \lambda \hat{b} \big) = \chi_{I J}\beta^I \beta^J  + \text{O}(\beta^3, \beta^2 \partial \beta) \,.
\end{equation}
This result dispels the concerns on the validity of the perturbative
$a$-theorem in $d=6$ raised by~\cite{Grinstein:2014xba}, where a different
function of the coupling constants was proposed as the monotonically
decreasing quantity.  As a direct consequence of the $a$-theorem we prove
the equivalence $\text{SFT}=\text{CFT}$ in our setup.


\section{The local Callan--Symanzik equation}
In this section we review briefly the local CS equation
formalism, which we will use to derive constraints on the RG flow.  The
local CS equation was first derived in the seminal work
\cite{Osborn:1991gm}. We refer the reader to \cite{Baume:2014rla} and
\cite{Jack:2013sha} for a detailed and thorough analysis of this technology
in four dimensions.

The RG flow is equivalent to a global rescaling of distances, which is
controlled by the properties of the trace of the stress-energy tensor, $T$.
In perturbation theory, the stress-energy tensor can be expanded in a basis
of operators of the CFT. Schematically,
\begin{equation}
\label{eq:Tbasis}
T \sim  \beta^I \mathcal{O}_I + S^A\lsp \nabla_\mu J^\mu_A
- \eta^a\lsp \nabla^2 \mathcal{O}_a
+ C^\alpha\lsp \nabla^2\nabla^2 \varphi_\alpha\,,
\end{equation}
where $\mathcal{O}_I$ are marginal scalar operators of dimension six,
$J^\mu_A$ are currents of dimension five generating an exact flavor
symmetry $G_F$ at the fixed point $\lambda^I=0$, while $\mathcal{O}_a$ and
$\varphi_\alpha$ are scalar operators of dimensions four and
two.\footnote{By the unitarity bound $\varphi_\alpha$ can only be free
fields satisfying $\nabla^2 \varphi_\alpha=0$ at the fixed point.}

For simplicity, in this work we will assume that the lower-dimensional
scalar operators $\mathcal{O}_a$ and $\varphi_\alpha$ are absent. It would
be interesting to include them in the future, also to further test
results in the perturbative $\phi^3$ theory \cite{Grinstein:2015ina}.

To express the response of the theory (\ref{eq:perturbedtheory}) under
local changes of the renormalization scale, it is necessary to turn on
sources for the renormalized operators in (\ref{eq:Tbasis}).  We lift the
theory to curved spacetime, such that the metric $g_{\mu \nu}(x)$ sources
the stress-energy tensor $T^{\mu \nu}$.  In addition, we promote the
couplings $\lambda^I(x)$ to spacetime dependent sources of the marginal
operators $\mathcal{O}_I$, and we introduce the background gauge fields
$A^A_\mu(x)$ sourcing the currents $J^\mu_A$. The $G_F$ symmetry is thus
gauged and $\lambda^I$ transform under the symmetry. The quantum effective
action then reads
\begin{equation}
\mathcal{W}[\mathcal J] = -i \log \int \mathcal{D} \Phi \,
e^{i S[\Phi, \mathcal{J}] }\,,
\end{equation}
where we collectively denote the sources as $\mathcal{J} \equiv (g^{\mu
\nu}(x), \lambda^I(x), A^\mu_A(x))$.

The connected correlation functions can be expressed as functional
derivatives with respect to the $\mathcal{J}$'s,
\begin{equation}
\frac{2}{\sqrt{-g}} \frac{\delta}{\delta g^{\mu \nu}(x)} \to
\left[ T_{\mu \nu}(x)\right]\,, \qquad
\frac{1}{\sqrt{-g}} \frac{\delta}{\delta \lambda^I(x)} \to
\left[ \mathcal{O}_I(x)\right]\,, \qquad
\frac{1}{\sqrt{-g}} \frac{\delta}{\delta A^A_\mu(x)} \to
\left[ J_A^\mu (x)\right]
\, ,
\end{equation}
where the square brackets denote the operator insertion inside a
renormalized correlation function.  For instance, the time-ordered
renormalized correlators of the scalar marginal operators are obtained as
\begin{equation}
\langle \mathbf {T} \left\{\mathcal{O}_{I_1}(x_1)\cdots
\mathcal{O}_{I_n}(x_n) \right\} \rangle =
\frac{(-i)^{n-1}}{\sqrt{-g(x_1)}\cdots\sqrt{-g(x_n)}}
\frac{\delta}{\delta \lambda^{I_1}(x_1)} \cdots
\frac{\delta}{\delta \lambda^{I_n}(x_n)} \mathcal{W}\,.
\end{equation}
To evaluate these correlation functions in the perturbed theory
(\ref{eq:perturbedtheory}) in flat space, one has to take $g^{\mu \nu}(x)
\to \eta^{\mu \nu}$, $\lambda^I(x) \to \lambda^I = {\rm const}$, $A^A_\mu(x) \to 0 $ after the variation.

To derive constraints on the RG flow we will consider the response of the
quantum effective action to a local change of the renormalization scale.
The local CS equation \cite{Osborn:1991gm} reads
\eqna{\label{eq:localCS}
\Delta_\sigma \mathcal{W} &\equiv \int \dsix x\sqrt{-g} \, \left( 2\lsp
\sigma g^{\mu \nu} \frac{\delta}{\delta g^{\mu \nu}} + \delta_\sigma
\lambda^I \frac{\delta}{\delta \lambda^I} + \delta_\sigma A^A_\mu \cdot
\frac{\delta}{\delta A^A_\mu} \right) \mathcal{W} = \int \dsix x\sqrt{-g}
\,\mathcal{A}_\sigma\,, \\ \delta_{\sigma} \lambda^I &=  -\sigma\lsp
\beta^I \,, \qquad \delta_{\sigma} A^A_\mu  =   -\sigma \rho^A_I\lsp
\nabla_\mu \lambda^I  + \partial_\mu \sigma\lsp S^A\,,}[]
where $\Delta_\sigma$ contains the most general terms allowed by covariance
and power counting, $\nabla$ is a gauge covariant derivative, and the
anomaly $\mathcal{A}_\sigma$ is a local functional of the sources, whose
form is constrained by diff-invariance and power counting.  The
Wess--Zumino consistency conditions,
\begin{equation}
\label{eq:cc}
\Delta_\sigma \mathcal{A}_{\sigma'} -
\Delta_{\sigma'} \mathcal{A}_{\sigma} = 0\,,
\end{equation}
expressing the commutativity of Weyl rescalings, impose further constraints
among the coefficients of the various terms that appear in $\cA_\sigma$. At
the fixed point, i.e.\ for $\lambda^I = {\rm const}$, $\beta^I= S^A = 0$ and $A^\mu_A= 0$,
$\mathcal{A}_\sigma$ reduces to the usual conformal
anomaly~\cite{Bonora:1985cq},
\begin{equation}
\label{eq:FPanomaly}
\mathcal{A}_\sigma = \sigma \, (-a \, E_6 + c_1 I_1 + c_2 I_2 + c_3 I_3)\,,
\end{equation}
up to six contributions (trivial anomalies) that can be eliminated by
adding local counterterms to the effective action.  In (\ref{eq:FPanomaly})
$E_6$ is the Euler term while $I_1$, $I_2$, $I_3$ are Weyl invariant
tensors. Their explicit form can be found in Appendix \ref{appAnom}. The
condition \eqref{eq:cc} at the fixed point imposes the vanishing of seven
other possible anomalies (analogous to the $R^2$ anomaly in $d=4$).

In the next section we are going to derive constraints on the RG flow
implied by the consistency conditions for the anomaly outside the fixed
point.

\section{Constraints on RG flows}
\label{sec:cc}
Consistency conditions that follow from the commutativity of Weyl
rescalings impose constraints among the various terms that appear in the
anomaly $\cA_\sigma$.  In $d=2,4$ these conditions were originally
considered in~\cite{Osborn:1991gm}, and were recently also studied in
detail in~\cite{Luty:2012ww, Fortin:2012hn, Jack:2013sha, Baume:2014rla},
and holographically in~\cite{Rajagopal:2015lpa, Kikuchi:2015hcm}. The
consistency conditions were also studied in supersymmetric theories
in~\cite{Auzzi:2015yia, Gillioz:2016ynj}. In $d=6$ they were first
considered in~\cite{Grinstein:2013cka}. Here we derive the consistency
conditions from the results of \eqref{eq:cc}, as obtained
in~\cite{Grinstein:2013cka}, and perform a detailed analysis of those.  We
find that some consistency conditions obtained in~\cite{Grinstein:2013cka}
were incomplete.

For the moment, we will neglect the contributions to equation
(\ref{eq:localCS}) related to the gauge fields $A^A_{\mu}$ sourcing the
currents $J_A^\mu$.  However, as will be shown in section
\ref{section:scaleconformal}, this will not change our conclusions.  The
complete form of $\mathcal{A}_\sigma$ can be found in Appendix
\ref{appAnom}.  After decomposing \eqref{eq:cc} in a linearly-independent
basis, it is possible to read off constraint equations for the anomaly
coefficients.  This is technically challenging, particularly due to
difficulties related to integration by parts and Bianchi
identities.\footnote{All our computations were performed in
\emph{Mathematica} using the package \texttt{xAct}, and details on the
derivation of the consistency conditions can be found in Appendix
\ref{ccderivation}.  Due to the large number of terms appearing in
\eqref{eq:cc} and related consistency conditions, we do not report most of
them in the text.  The interested reader can find them in a separate
\emph{Mathematica} file attached to the submission.} The consistency
conditions obtained here were checked at two loops in the $\phi^3$ theory
against the results of~\cite{Grinstein:2015ina}.\foot{To extend the check
to higher loops it will be necessary to include the effects of the
operators of dimension two and four.} We have also checked that they are
satisfied by the general form of the trace anomaly on the conformal
manifold as constucted in~\cite{Osborn:2015rna}.

In this work we exploit all constraints imposed on anomaly coefficients
with up to two indices. This requires us to decompose the consistency
conditions and isolate the ones that stem from terms involving up to two
couplings $\lambda$. For example, we are interested in the consistency
condition arising from contributions to the left-hand side of \eqref{eq:cc}
proportional to $(\sigma\lsp\partial_\mu\sigma^\prime
-\sigma^\prime\lsp\partial_\mu\sigma) \nabla^2\lambda^I\lsp\partial^\mu
\nabla^2\lambda^J$, but not in the one arising from contributions
proportional to $(\sigma\lsp\partial_\mu\sigma^\prime
-\sigma^\prime\lsp\partial_\mu\sigma)\lsp\partial^\mu\lambda^I\lsp
\nabla^2\lambda^J\lsp \nabla^2\lambda^K$.

A particularly important equation contained in \eqref{eq:cc}
is obtained from terms proportional to $( \sigma \lsp\partial_\mu \sigma' -
\sigma'\lsp \partial_\mu \sigma) H_1{}^{\mu \nu} \partial_\nu
\lambda^I$, where $H_1{}^{\mu \nu}$ is a generalization of the Einstein
tensor in $d=6$ \cite{Lovelock:1971yv} (see \eqref{BFour} for its explicit
form), namely
\begin{equation}
\label{eq:atildeeq}
\partial_I \check{a} = \tfrac{1}{6}\lsp\mathcal{H}_{I J}\beta^J +
\tfrac{1}{6}\lsp\mathcal{H}_I\,,
\end{equation}
where
\eqn{\begin{gathered}
\check{a}=a +\tfrac{1}{6\lsp} b_{1} -\tfrac{1}{90}\lsp b_3
+\tfrac{1}{6}\lsp b_{11} +\tfrac{1}{12}\lsp \mathcal{A}_J \beta^J
+ \tfrac{1}{6}\lsp \mathcal{H}^1_J \beta^J\,,\\
\mathcal{H}_I= -\mathcal{H}^5_{I} - \tfrac{1}{2}\lsp
\mathcal{H}^6_{I} - \tfrac{1}{2}\lsp \mathcal{I}^7_{I}\,,\qquad
\mathcal{H}_{I J}= \tfrac{1}{4}\lsp \mathcal{A}_{JI}
+ \mathcal{H}^1_{I J} + \partial_I \mathcal{A}_J + \partial_{[I}
\mathcal{H}^1_{J]}\,,
\end{gathered}}[anomcoeffI]
with the definition
\eqn{\partial_{[I}X_{J]}=\partial_IX_J-\partial_JX_I\,.}[antisym]
All tensors appearing above are local functions of the couplings, and their
definition can be found in Appendix~\ref{appAnom}.  Use of the consistency
condition arising from $(\sigma\lsp\nabla^\mu\nabla^\nu\partial^\rho
\sigma^\prime- \sigma^\prime\lsp \nabla^\mu\nabla^\nu\partial^\rho \sigma)
\nabla_\mu\nabla_\nu \partial_\rho\lambda^I$ allows us to put
\eqref{eq:atildeeq} in the form
\begin{equation}
\label{eq:atildeeq2}
\partial_I \tilde{a} = \tfrac{1}{6}(\mathcal{H}^1_{I J}-\tfrac14\lsp
\hat{\cA}_{IJ}^{\prime\prime})\beta^J
+\tfrac16\lsp\partial_{[I}\cH_{J]}^1\beta^J
-\tfrac{1}{12}\lsp\cI_I^7\,,\qquad
\tilde{a}= a+\tfrac16\lsp b_1-\tfrac{1}{90}\lsp
b_3+\tfrac16\lsp\cH_I^1\beta^I\,,
\end{equation}
which contains fewer anomaly coefficients than \eqref{eq:atildeeq} with
\anomcoeffI.  Unlike in the two and four-dimensional cases,
(\ref{eq:atildeeq2}) does not present itself in the form of
(\ref{eq:gradientflow}), due to the presence of the vector anomaly
$\mathcal{I}^7_I$. Notice that this contribution was missed in
\cite{Grinstein:2013cka}, which led to consider $\tilde{a}$ as the
candidate for a monotonically-decreasing function in
\cite{Grinstein:2014xba}.  However, $\tilde{a}$ cannot be such a candidate,
even more so because it is scheme-dependent\footnote{In this paper, by
``scheme-dependent'' quantities we mean those which change under the
addition of purely background-dependent counterterms to the effective
action.} at order $\beta$.\footnote{For example, the addition of a term $\int\dsix
x\sqrt{\gamma}\,X_I\lsp\partial_\mu\lambda^I\lsp\nabla_\nu H_4^{\mu\nu}$ in
$\mathcal{W}[\mathcal{J}]$, with $X_I$ arbitrary, induces, among others,
the shifts $\cI_{I}^7\to\cI_I^7+\mathscr{L}_\beta X_I$, where
$\mathscr{L}_\beta$ is the Lie derivative along the beta function, and
$\cH^1_I\to\cH_I^1-\frac{1}{2}\lsp X_I$. The shift of $\cH_{I}^1$ affects
$\tilde{a}$ at order $\beta$.}

In this work we consider linear combinations of the consistency conditions
in order to find all independent equations having the form of
\eqref{eq:gradientflow}.  Most importantly, we find the
equation\footnote{The linear combination of the consistency conditions
leading to \eqref{eq:gradientflow1} is explicitly reported in the
\emph{Mathematica} file attached to the submission.}
\begin{equation}
\label{eq:gradientflow1}
\partial_I \hat{a} = (\chi_{I J} + \xi_{I J})\beta^J \,,
\end{equation}
where
\begin{equation}
\begin{split}
\hat{a} &= a - \tfrac{5}{6}\lsp  b_1 +\tfrac{1}{10}\lsp b_2 + \tfrac{1}{45}\lsp b_3 + \tfrac{1}{10}\lsp b_4
\\ &\quad
+ \left(\tfrac{1}{10}\lsp \cB_I +\tfrac{1}{24}\lsp \cC_I + \tfrac{1}{20}\lsp \cE_I +\tfrac{1}{24}\lsp\cF_I
+\tfrac{1}{6}\lsp\cH^1_I +\tfrac{1}{20}\lsp\cH^2_I +\tfrac{1}{12}\lsp\cH^3_I +\tfrac{1}{8}\lsp\cH^4_I -\tfrac{1}{40}\lsp\cH^6_I \right)\beta^I\,,
\\ \chi_{IJ} &= \tfrac{1}{20}\lsp\partial_{(I} \cB_{J)} -\tfrac{1}{40}\lsp \hat{\cB}'_{IJ} + \tfrac{1}{48}\lsp\hat{\cC}'_{IJ}  +\tfrac{1}{20}\lsp\hat{\cE}_{(I J)}
+ \tfrac{1}{24}\lsp\cF_{(I J)}
+\tfrac{1}{6}\lsp\cH^1_{IJ}
\\ &\quad
+\tfrac{1}{20}\lsp\cH^2_{IJ} + \tfrac{1}{12}\lsp\cH^3_{IJ}
+\tfrac{1}{8}\lsp\cH^4_{IJ} -\tfrac{1}{40}\lsp\cH^6_{IJ}\,,
\\ \xi_{I J} &= \tfrac{1}{20}\lsp \partial_{[I}\cB_{J]}
+\tfrac{1}{48}\lsp\cC_{[I J]}  +
\tfrac{1}{40}\lsp \hat{\cE}_{[I J]} + \tfrac{1}{48}\lsp \cF_{[I J]} +
\tfrac{1}{48}\lsp \cF'_{[I J]}
\\ &\quad + \tfrac{1}{6}\lsp \partial_{[I} \cH^1_{J]} +
\tfrac{1}{20}\lsp\partial_{[I} \cH^2_{J]} +
\tfrac{1}{12}\lsp\partial_{[I} \cH^3_{J]} + \tfrac{1}{8}\lsp
\partial_{[I}\cH^4_{J]} - \tfrac{1}{40}\lsp \partial_{[I}
\cH^6_{J]}\,,
\end{split}
\end{equation}
and we use \antisym and
\eqn{\partial_{(I}X_{J)}=\partial_IX_J+\partial_JX_I\,,\qquad
X_{(IJ)}=X_{IJ}+X_{JI}\,,\qquad
X_{[IJ]}=X_{IJ}-X_{JI}\,.}[]
$\hat{a}$ equals $a$ at the fixed point, for the anomalies $b_{1,\ldots,7}$
are all proportional to $\beta$.  $\chi_{I J}$ and $\xi_{I J}$ are
symmetric and antisymmetric tensors, respectively.\footnote{Using the
consistency conditions we have checked that $\xi_{I J}$ cannot be written
as $\partial_{[I}X_{J]}$ for some vector $X_J$.} Note that, by virtue of
equation (\ref{eq:gradientflow1}), $\hat{a}$ is scheme independent at order
$\beta$, while $\chi_{I J}$ and $\xi_{I J}$ are scheme-independent at order
$\beta^0$, i.e. they are not affected to that order by adding local
counterterms to the effective action.

Now we can show that the metric $\chi_{IJ}$ in \eqref{eq:gradientflow1} is
positive-definite. Indeed, consider the RG derivative of the two-point
correlator of the marginal operators $\langle \cO_I(x)\lsp \cO_J(0)
\rangle$ (in Euclidean signature).  First notice that, since lengths are
entirely controlled by $g_{\mu \nu}$, this operation can be expressed as a
Weyl rescaling,
\begin{equation}
\mu \frac{\partial}{\partial \mu} \mathcal{W} = - 2 \int \dsix x
\,g^{\mu \nu} \frac{\delta}{\delta g^{\mu \nu}} \mathcal{W}\,.
\end{equation}
Then, neglecting terms involving the $\beta$ function,
\eqna{\mu\frac{\partial}{\partial\mu}\vev{\cO_I(x)\lsp\cO_J(0)} &= - \int
\dsix y\frac{1}{\sqrt{-g(x)}\sqrt{-g(0)}}  \left(2 \lsp g^{\mu \nu}(y)
\frac{\delta}{\delta g^{\mu \nu}(y)} \right)
\frac{\delta}{\delta \lambda^{I}(x)}
\frac{\delta}{\delta \lambda^{J}(0)} \mathcal{W}
\\ &= - \int \dsix y\sqrt{-g(y)}\, \frac{1}{\sqrt{-g(x)}}\frac{\delta}{\delta
\lambda^{I}(x)}\frac{1}{\sqrt{-g(0)}}\frac{\delta}{\delta \lambda^{J}(0)} \mathcal{A}_{\sigma=1}\\
& = g_{IJ}(\partial^2)^3\delta^{(6)}(x)\,,}[]
where in the last line we go to flat space, $\delta^{(6)}(x)$ is the
six-dimensional delta function, and $g_{I J}$ is evaluated via the anomaly
in Appendix~\ref{appAnom},
\begin{equation}
g_{I J} = -\partial_{(I} \cA_{J)} -\hat{\cA}_{(I J)} + \hat{\cA}'_{I J}+ \hat{\cA}''_{I J}\,.
\end{equation}
It can be shown that $g_{IJ}$ is proportional to the Zamolodchikov metric
and is thus positive-definite by unitarity \cite{Osborn:2015rna}.
Furthermore, the consistency conditions relate the tensors $\chi_{I J}$ and
$g_{IJ}$ via
\begin{equation}
\chi_{I J} = \tfrac{1}{6}\lsp g_{I J} + \text{O}(\beta, \partial \beta)\,.
\end{equation}
With this result, and upon contracting equation (\ref{eq:gradientflow1})
with $\beta^I$, we get the desired monotonicity constraint in perturbation
theory for $\hat{a}$,
\begin{equation}
  \mu \frac{d }{d \mu} \hat{a} = \lsp \chi_{I J}\beta^I \beta^J \geq 0\,,
\end{equation}
where the inequality is saturated only if $\beta^I=0$. This proves the
$a$-theorem in perturbation theory (in theories with no relevant scalar
operators of dimension two and four).

Additionally, we find another, independent equation of the form\footnote
{The linear combination of the consistency conditions leading to
\eqref{eq:gradientflow2} is explicitly reported in the \emph{Mathematica}
file attached to the submission.}
\eqn{\partial_I \hat{b}= (\chi'_{I J} + \xi'_{I J})\beta^J\,,
}[eq:gradientflow2]
where
\begin{equation}
\begin{split}
\hat{b} &= 4\lsp b_1 -\tfrac{4}{5}\lsp b_2 -\tfrac{4}{15}\lsp b_3
-\tfrac{4}{5}\lsp b_4 -\left(\tfrac{4}{5}\lsp \cB_I +\tfrac{1}{2}\lsp \cC_I
+\tfrac{2}{5}\lsp \cE_I
+\tfrac{2}{5}\lsp \cH^2_I +\tfrac{2}{3}\lsp\cH^3_I +\tfrac{2}{3}\lsp
\cH^4_I -\tfrac{1}{5}\lsp \cH^6_I \right)\beta^I\,,
\\ \chi'_{I J} &= - \tfrac{2}{5}\lsp \partial_{(I}
\cB_{J)}+\tfrac{1}{3}\lsp\hat{\cA}''_{I J} + \tfrac{1}{5}\lsp \hat{\cB}'_{I J} - \tfrac{1}{6}\lsp\hat{\cC}'_{I J} -\tfrac{1}{5}\lsp \hat{\cE}_{(I J)}
- \tfrac{2}{5}\lsp \cH^2_{I J} - \tfrac{2}{3}\lsp \cH^3_{I J}
-\tfrac{2}{3}\lsp \cH^4_{I J} + \tfrac{1}{5}\lsp \cH^6_{I J}\,,
\\ \xi'_{I J} &= -\tfrac{2}{5}\lsp \partial_{[I} \cB_{J]} - \tfrac{1}{5}\lsp \hat{\cE}_{[I J]} -\tfrac{2}{5}\lsp \partial_{[I} \cH^2_{J]} -\tfrac{2}{3}\lsp \partial_{[I}\cH^3_{J]} -\tfrac{2}{3}\lsp\partial_{[I}\cH^4_{J]}
+ \tfrac{1}{5}\lsp \partial_{[I} \cH^6_{J]}\,.
\end{split}
\end{equation}
$\hat{b}$ is of order $\beta$ and so vanishes at fixed points, and
$\chi'_{I J}$, $\xi'_{I J}$ are symmetric and antisymmetric respectively.
The existence of the metric $\chi'_{I J}$ is related to the fact that in
$d=6$ there are three rank-two conformally covariant operators one can
define on the conformal manifold~\cite{Osborn:2015rna}, corresponding to
just as many scheme-independent rank-two tensors at the fixed point.  This
is in contrast with the two- and four-dimensional cases where there is only
a unique rank-two tensor related to the Zamolodchikov metric.
Nevertheless, we found that the consistency conditions impose an
orthogonality constraint on $\chi'_{I J}$,
\begin{equation}
\label{eq:orthogonality}
\chi'_{I J} \beta^J = \text{O}(\beta^2, \beta \partial \beta) \,,
\end{equation}
even though, in general, $\chi'_{I J}$ does not vanish at fixed points.
Equations (\ref{eq:gradientflow1}), (\ref{eq:gradientflow2}), (\ref{eq:orthogonality})
 imply that there exists a one-parameter family of
 monotonically decreasing functions at leading order in perturbation theory,
\begin{equation}
\mu \frac{d}{d \mu} \big(\hat{a} + \lambda \hat{b}\big)
= \tfrac{1}{6}\lsp g_{I J} \beta^I \beta^J + \text{O}(\beta^3, \beta^2 \partial \beta) \,.
\end{equation}

\section{Scale versus conformal invariance}
\label{section:scaleconformal}

We could ask whether the theory (\ref{eq:perturbedtheory}) can flow to a
nearby scale invariant field theory without conformal invariance.  This
question becomes nontrivial in presence of dimension five currents, as we
see from equation (\ref{eq:Tbasis}).  Indeed, at the fixed point it is
$\beta^I=0$, and $T$ has the operatorial form
\begin{equation}
T \sim S^A \nabla_\mu J^\mu_A \equiv \nabla_\mu V^{\mu}\,,
\end{equation}
where $V^{\mu}$ is the so-called virial current. If $S^A \ne 0$ the theory
is scale but not conformally invariant, for $T$ is a total divergence.

Before proceeding, it is useful to rewrite the Weyl operator in a more
convenient form.  By making use of the Ward identities for the $G_F$
symmetries represented by the broken generators $T^A$, it possible to
redefine the Weyl operator in (\ref{eq:localCS}) to encapsulate both a Weyl
rescaling and a $G_F$ transformation~\cite{Osborn:1991gm, Fortin:2012hn,
Jack:2013sha, Baume:2014rla},
\eqna{\label{eq:localCSrotated}
\Delta'_\sigma \mathcal{W} &\equiv \int \dsix x\sqrt{-g} \,
\sigma\left( 2 \lsp g^{\mu \nu} \frac{\delta}{\delta g^{\mu \nu}}
- B^I\frac{\delta}{\delta \lambda^I} -P^A_I\lsp \nabla_\mu \lambda^I \cdot \frac{\delta}{\delta A^A_\mu} \right) \mathcal{W}
= \int \dsix x\sqrt{-g}\, \mathcal{A}_\sigma\,,
\\ B^I &= \beta^I - (S^A T_A \lambda)^I \,, \qquad P^A_I = \rho^A_I +
\partial_I S^A\,,}[]
with the constraint $B^I P^A_I = 0$ due to the
commutativity of Weyl rescaling, $[\Delta'_{\sigma}, \Delta'_{\sigma'}] = 0$.
In this parametrization a scale invariant field theory corresponds to $B^I
= -(S^A T_A \lambda)^I$, while a conformal invariant field theory to $B^I =
0$.

Now, let us generalize the equation (\ref{eq:gradientflow1}) in the
presence of dimension five currents.  By covariance, at leading order in
$B^I$ it takes the form
\begin{equation}
\label{eq:gradientflowWithGauge}
\partial_I \hat{a} = (\chi_{I J} + \xi_{I J})B^J + P^A_I f_A\,,
\end{equation}
where $f_A$ is an generic combination of anomaly coefficients of terms
involving the gauge fields $A^A_\mu$.  Upon contracting
(\ref{eq:gradientflowWithGauge}) with $B^I$ and using the condition $B^I P^A_I = 0$ we get
\begin{equation}
\label{eq:floweqWithGauge}
B^I \partial_I \hat{a} = \mu \frac{d}{d \mu} \hat{a} = \tfrac{1}{6}\lsp g_{I J} B^I B^J \geq 0\,.
\end{equation}
Therefore, we reach the same conclusion we found in section \ref{sec:cc}.
Furthermore, suppose we are in a scale invariant field theory. Then,
$G_F$-invariance of $\hat{a}$ implies that $B^I \partial_I \hat{a} =
-(S^A T_A \lambda)^I \partial_I \hat{a} = 0$,
so that (\ref{eq:floweqWithGauge}) gives
\begin{equation}
g_{I J} B^I B^J = 0\,.
\end{equation}
Due to the positive-definiteness of $g_{I J}$ this can only be true for
$B^I=0$. This proves that scale invariance implies conformal invariance in
our setup, in analogy with the four-dimensional case~\cite{Fortin:2012hn,
Luty:2012ww}. Our proof here follows the logic of~\cite{Fortin:2012hn}.

\section{Conclusions}
In this work we studied the properties of RG flows originating from
marginal deformations to unitary conformal field theories in six
dimensions.  For simplicity, we restricted the analysis to a class of CFTs
where relevant scalar operators of dimension two and four are absent.  Even
though we work in perturbation theory, the UV CFT can in general be
strongly coupled and may not admit a Lagrangian description.

The results obtained here can be summarized as follows:
\begin{itemize}
\item We derived all the consistency conditions with up to two powers of
  the coupling outside the fixed point.  We solved those to find all the
  constraints among the anomaly coefficients which can be put in the form
  of a flow equation.

\item We identified a one-parameter family of
  scheme-independent functions of the coupling constants of the theory,
  $\hat{a}+\lambda \hat{b}$ with $\lambda \in \mathbb{R}$, equal to the
  $a$-anomaly coefficient plus $\text{O}(\beta)$ corrections, which flow
  monotonically in the proximity of a fixed point thanks to unitarity.
  There is no parameter $\lambda$ for which the combination
  $\hat{a}+\lambda \hat{b}$, agrees with the quantity analyzed in
  \cite{Grinstein:2014xba} in the context of $\phi^3$ theory,
  therefore we dispel the doubts cast on the perturbative $a$-theorem
  in six dimensions.

\item As a direct consequence of the $a$-theorem we proved, using standard
  arguments, that scale implies conformal invariance in our setup.
\end{itemize}

The dynamics of perturbative QFTs in six dimensions appears structurally
different with respect to the four-dimensional case, due to the presence of
multiple scheme-independent rank two tensors at the fixed point.
Nevertheless, we were able to find a class of physical quantity whose RG flow is
governed uniquely by the positive definite Zamolodchikov metric.
We presume that extending our argument beyond perturbation theory would
single out the monotonically-decreasing function in the one-parameter
family that we found.

In the future, it will be interesting to extend our results in the presence
of scalar operators of dimension two and four. First, that could highlight
possible differences with the lower spacetime dimensional cases, where
relevant operators do not affect the monotonicity
constraints~\cite{Osborn:1991gm, Jack:2013sha, Baume:2014rla}.\footnote{In
four dimensions that is made clear by the argument employing the on-shell
dilaton amplitude, which is manifestly insensitive to those
effects~\cite{Baume:2014rla}.} Second, that will be necessary to test our
results in the $\phi^3$ theory, which is the only perturbatively calculable
theory in six dimensions. It should be straightforward to generalize our
computations to include those contributions, with the only difficulties
arising due to the proliferation of terms in the anomaly functional and in
the Weyl operator.

It would also be of interest to analyze $\hat{a}$ and $\hat{b}$ to
higher-loop orders in $\phi^3$ theory with the use of the consistency
conditions, along the lines of~\cite{Gracey:2015fia}. The effects of
dimension two and four operators as described in the previous paragraph may
be necessary for such an analysis.

The question stands whether the $a$-theorem and the equivalence of scale
and conformal invariance is valid beyond perturbation theory in six
dimensions. So far no counterexamples are known.  In four dimensions,
certain dilaton scattering amplitudes provide a powerful tool to address
these questions~\cite{Komargodski:2011vj, Dymarsky:2013pqa}.  Attempts were
made to use dilaton scattering amplitudes \cite{Elvang:2012st} in six
dimensions, but it is not clear what the right approach would be.

\ack{For our computations we have relied heavily on \emph{Mathematica} and
the package \href{http://www.xact.es/}{\texttt{xAct}}. We thank Riccardo
Rattazzi for his valuable input and suggestions throughout this project. We
are grateful to Ben Grinstein, Hugh Osborn, and Riccardo Rattazzi for
comments on the manuscript. AS would like to thank Brian Henning for
helpful discussions.  The research of AS is supported in part by the
National Science Foundation under Grant No.~1350180. The research of DS is
supported by a grant from the European Research Council under the European
Union's Seventh Framework Programme (FP 2007-2013) ERC Grant Agreements
No.\ 279972 ``NPFlavour''.  The work of LGV is supported by the Swiss
National Science Foundation under grant 200020-150060.}

\begin{appendices}

\newsec{Derivation of the consistency conditions}[ccderivation]

In this work, in order to derive the consistency conditions it was
necessary to write the variation \ref{eq:cc} in a linearly independent
basis.  This was technically nontrivial due to the large number of terms
($\sim \text{O}(100)$) and redundancies related to integration by parts.
Our approach is outlined in this appendix. First, by integrating by
parts, we took all the derivatives off either $\sigma$ or $\sigma'$. As a
result, we ended up with terms such as
\begin{equation}
\label{eq:anomalyVarAfterIBP}
(\sigma\lsp \partial_\mu \sigma' - \sigma'\lsp \partial_\mu \sigma)\,
f_I(\lambda)\lsp \partial^\mu \lambda^I \lsp R^2
\,, \qquad (\sigma\lsp \nabla_\mu \partial_\nu \sigma' -
\sigma'\lsp \nabla_\mu \partial_\nu \sigma) \,f(\lambda)\lsp
H_1^{\mu \nu}\,.
\end{equation}
However, there are still redundancies related to antisymmetrization with
respect to $\sigma, \sigma'$.  For example, consider the trivial equation
\begin{equation}
(\partial_\mu \sigma\lsp \partial_\nu \sigma' - \partial_\mu \sigma'\lsp
\partial_\nu \sigma)\, f(\lambda)\lsp H_1^{\mu \nu} = 0\,,
\end{equation}
where $H_1^{\mu \nu}$ is symmetric. Upon integrating by parts and writing
this equation in the same basis as (\ref{eq:anomalyVarAfterIBP}), we get
\begin{equation}
(\sigma\lsp \nabla_\mu \partial_\nu \sigma'- \sigma'\lsp \nabla_\mu
\partial_\nu \sigma)\, f(\lambda) \lsp H_1^{\mu \nu}
+(\sigma\lsp \partial_\nu \sigma' -\sigma'\lsp \partial_\nu \sigma)\,
\partial_I f(\lambda)\lsp \partial_\mu \lambda^I\lsp  H_1^{\mu \nu} = 0\,,
\end{equation}
since $\nabla_\mu H_1^{\mu \nu} = 0$ in this example. This allows to
eliminate the second term in (\ref{eq:anomalyVarAfterIBP}).  Similarly one
can get rid of all the terms with an even number of derivatives on
$\sigma$, $\sigma'$. This prescription fixes unambiguously a complete basis
for \eqref{eq:cc}.

\newsec{Conventions and basis for the anomaly}[appAnom]
We define the Riemann tensor via
\eqn{[\nabla_{\mu},\nabla_{\nu}] A^{\rho} =
R^\rho{\!}_{\sigma\mu\nu} A^{\sigma} \,,}[mswConvention]
and the Ricci tensor and Ricci scalar as
$R_{\mu\nu}=R^{\rho}{\!}_{\mu\rho\nu}$ and $R=g^{\mu\nu}R_{\mu\nu}$.
The Einstein tensor is defined in $d\geq2$ by
\eqn{G_{\mu\nu}=\tfrac{2}{d-2}(R_{\mu\nu}-\tfrac12\lsp
g_{\mu\nu}R)\,,}[]
while the Weyl tensor is defined in $d\geq3$ by
\eqn{W_{\mu\nu\rho\sigma}=R_{\mu\nu\rho\sigma}+
\tfrac{2}{d-2}(g_{\mu[\sigma}R_{\rho]\nu}+
g_{\nu[\rho}R_{\sigma]\mu})+
\tfrac{2}{(d-1)(d-2)}g_{\mu[\rho}g_{\sigma]\nu}R\,.}[]
At dimension four we consider the tensors
\eqn{\begin{gathered}
E_4=\tfrac{2}{(d-2)(d-3)}(R^{\mu\nu\rho\sigma}R_{\mu\nu\rho\sigma}
  -4R^{\mu\nu}R_{\mu\nu}+R^2)\,,\qquad
I=W^{\mu\nu\rho\sigma}W_{\mu\nu\rho\sigma}\,,\\
H_{1\llsp\mu\nu}=\tfrac{(d-2)(d-3)}{2}E_4 \lsp g_{\mu\nu}
  -4(d-1)H_{2\mu\nu}
  +8H_{3\mu\nu}+8H_{4\mu\nu}
  -4R^{\rho\sigma\tau}{\!}_{\mu}R_{\rho\sigma\tau\nu}\,,\\
H_{2\llsp\mu\nu}=\tfrac{1}{d-1}RR_{\mu\nu}\,,\qquad
H_{3\llsp\mu\nu}=R_{\mu}{\!}^{\rho}R_{\rho\nu}\,,\qquad
H_{4\llsp\mu\nu}=R^{\rho\sigma}R_{\rho\mu\sigma\nu}\,,\\
H_{5\llsp\mu\nu}=\nabla^2 R_{\mu\nu}\,,\qquad
H_{6\llsp\mu\nu}=\tfrac{1}{d-1}\nabla_\mu\partial_\nu R\,.
\end{gathered}}[BFour]
A complete basis of scalar dimension-six curvature terms consists of
\cite{Bonora:1985cq}
\eqn{\begin{gathered}
K_1=R^3\,,\qquad
K_2=RR^{\mu\nu}R_{\mu\nu}\,,\qquad
K_3=RR^{\mu\nu\rho\sigma}R_{\mu\nu\rho\sigma}\,,\qquad
K_4=R^{\mu\nu}R_{\nu\rho}R^\rho_{\hphantom{\rho}\!\mu}\,,\\
K_5=R^{\mu\nu}R^{\rho\sigma}R_{\mu\rho\sigma\nu}\,,\qquad
K_6=R^{\mu\nu}R_{\mu\rho\sigma\tau}
  R_{\nu}^{\smash{\hphantom{\nu}\rho\sigma\tau}}\,,\qquad
K_7=R^{\mu\nu\rho\sigma}R_{\rho\sigma\tau\omega}
R^{\tau\omega}{\!}_{\mu\nu}\,,\\
K_8=R^{\mu\nu\rho\sigma}R_{\tau\nu\rho\omega}
R_{\mu}{\!}^{\tau\omega}{\!}_\sigma\,,\qquad
K_9=R\,\nabla^2 R\,,\qquad
K_{10}=R^{\mu\nu}\,\nabla^2 R_{\mu\nu}\,,\qquad
K_{11}=R^{\mu\nu\rho\sigma}\,\nabla^2 R_{\mu\nu\rho\sigma}\,,\\
K_{12}=R^{\mu\nu}\nabla_\mu\partial_\nu R\,,\qquad
K_{13}=\nabla^\mu R^{\nu\rho}\lsp\nabla_{\mu}R_{\nu\rho}\,,\qquad
K_{14}=\nabla^\mu R^{\nu\rho}\lsp\nabla_{\nu}R_{\mu\rho}\,,\\
K_{15}=\nabla^\mu R^{\nu\rho\sigma\tau} \lsp
\nabla_\mu R_{\nu\rho\sigma\tau}\,,\qquad
K_{16}=\nabla^2 R^2\,,\qquad
K_{17}=(\nabla^2)^2 R\,.
\end{gathered}}
In $d=6$ a convenient basis is given by
\eqna{I_1&=\tfrac{19}{800}\llsp K_1 - \tfrac{57}{160}\llsp K_2 + \tfrac{3}{40}\llsp K_3 +
\tfrac{7}{16}\llsp K_4 - \tfrac{9}{8}\llsp K_5 - \tfrac{3}{4}\llsp K_6 +
\llsp K_8\,,\\
I_2&=\tfrac{9}{200}\llsp K_1 - \tfrac{27}{40}\llsp K_2 + \tfrac{3}{10}\llsp K_3 +
\tfrac{5}{4}\llsp K_4 - \tfrac{3}{2}\llsp K_5 - 3\llsp K_6 + \llsp K_7\,,\\
I_3&=-\tfrac{11}{50}\llsp K_1 + \tfrac{27}{10}\llsp K_2 - \tfrac{6}{5}\llsp K_3 - \llsp K_4 + 6\llsp K_5+
2\llsp K_7 - 8\llsp K_8\\&\hspace{3cm} + \tfrac{3}{5}\llsp K_9 - 6\llsp K_{10} + 6\llsp K_{11} + 3\llsp K_{13}
- 6\llsp K_{14} + 3\llsp K_{15}\,,\\
E_6&=\llsp K_1 - 12\llsp K_2 + 3\llsp K_3 + 16\llsp K_4 - 24\llsp K_5 - 24\llsp K_6 + 4\llsp K_7 +
8\llsp K_8\,,\\
J_1&=6\llsp K_6 - 3\llsp K_7 + 12\llsp K_8 + \llsp K_{10} - 7\llsp K_{11} - 11\llsp K_{13} + 12\llsp K_{14} -
  4\llsp K_{15}\,,\\
J_2&=-\tfrac15 \llsp K_9 + \llsp K_{10} + \tfrac25 \llsp K_{12} + \llsp K_{13}\,,\qquad
J_3=\llsp K_4 + \llsp K_5 - \tfrac{3}{20}\llsp K_9 + \tfrac45 \llsp K_{12}
  + \llsp K_{14}\,,\\
J_4&=-\tfrac15 \llsp K_9 + \llsp K_{11} + \tfrac25 \llsp K_{12} + \llsp K_{15}\,,\qquad
J_5=\llsp K_{16}\,,\qquad
J_6=\llsp K_{17}\,,\\
L_{1}&=-\tfrac{1}{30}\llsp K_1+\tfrac14 \llsp K_2-\llsp K_6\,,\qquad
L_{2}=-\tfrac{1}{100}\llsp K_1+\tfrac{1}{20}\llsp K_2\,,\\
L_{3}&=-\tfrac{37}{6000}\llsp K_1+\tfrac{7}{150}\llsp K_2 -\tfrac{1}{75}\llsp K_3
+\tfrac{1}{10}\llsp K_5+\tfrac{1}{15}\llsp K_6\,,\qquad
L_{4}=-\tfrac{1}{150}\llsp K_1+\tfrac{1}{20}\llsp K_3\,,\\
L_{5}&=\tfrac{1}{30}\llsp K_1\,,\qquad
L_{6}=-\tfrac{1}{300}\llsp K_1+\tfrac{1}{20}\llsp K_9\,,\qquad
L_{7}=\llsp K_{15}\,,}[]
where the first three transform covariantly under Weyl variations, and
$E_6$ is the Euler term in $d=6$.  The $J$'s are trivial anomalies in a
six-dimensional CFT defined in curved space, and the first six $L$'s are
constructed based on the relation $\delta_\sigma\int \dsix x\sqrt{-g}\,
L_{1,\ldots,6}=\int \dsix x\sqrt{-g}\,\sigma J_{1,\ldots,6}$.

In six spacetime dimensions there are ninety four independent terms that
can contribute to the anomaly~\cite{Grinstein:2013cka}. In general, we can
write
\eqn{ \int \dsix x\sqrt{-g}\,\mathcal{A}_\sigma=
\sum_{p=1}^{65}\int \dsix x\sqrt{-g}\,\sigma\st_p
+ \sum_{q=1}^{30}\int \dsix
x\sqrt{-g}\,\partial_\mu\sigma\,\sz^\mu_q,}[fullAnom]
where $\st_p$ and $\sz_q^\mu$ are dimension-six and dimension-five
terms respectively, that can involve curvatures as well as derivatives on
the couplings $\lambda^I$. In writing down the various terms below, we neglect
total derivatives.

If only curvatures are included, then we have the terms
\eqn{\begin{gathered}
\st_1=-c_1I_1\,,\qquad
\st_2=-c_2I_2\,,\qquad
\st_3=-c_3I_3\,,\qquad
\st_4=-aE_6\,,\qquad
\st_{5,\ldots,11}=-b_{1,\ldots,7}L_{1,\ldots,7}\,.
\end{gathered}}[]
We also have the terms
\eqn{\begin{gathered}
\sz_1^\mu=-b_8\,\partial^\mu E_4\,,\qquad
\sz_2^\mu=-b_9\,\partial^\mu I\,,\qquad
\sz_3^\mu=-\tfrac{1}{25}b_{10}\,R\,\partial^\mu R\,,\\
\sz_4^\mu=-\tfrac15 b_{11}\,\partial^\mu\nabla^2 R\,,\qquad
\sz_{5,6,7}^\mu=-b_{12,13,14}\,\nabla_\nu H_{2,3,4}^{\mu\nu}\,.
\end{gathered}}[insteadJ]
Actually, the terms in \insteadJ overcomplete the basis of trivial
anomalies.  This is because there are six trivial anomalies, but
seven terms in \insteadJ.  If we integrate the \insteadJ terms by parts,
then we may require that $\nabla_\mu\sz_{1,\ldots,7}^\mu$ do not affect the
coefficients of $L_{1,\ldots,7}$. This forces us to impose

\eqn{b_{13}=-\frac{24}{\dtwo-5d+6}b_8+\frac{4(d-6)}{d-2}b_9
-\frac{5}{d-1}b_{12}.}[choice]
With \choice it is guaranteed that $L_{1,\ldots,7}$ are vanishing
anomalies, and we also see that the coefficients of $E_6,I_{1,2,3}$ are
unaffected by $\nabla_\mu\sz_{1,\ldots,7}^\mu$.  Thus, with the condition
\choice the terms $\sz_{1,\ldots,7}^\mu$ substitute exactly the trivial
anomalies $J_{1,\ldots,6}$.

Next, we have
\eqn{\begin{gathered}
\st_{12}=\cI_I^1\,\partial_\mu \lambda^I\,\partial^\mu E_4\,,\qquad
\st_{13}=\cI_I^2\,\partial_\mu \lambda^I\,\partial^\mu I\,,\qquad
\st_{14}=\tfrac{1}{25}\cI_I^3\,\partial_\mu \lambda^I\,R\,\partial^\mu R\,,
\\
\st_{15}=\tfrac{1}{5}\cI_I^4\,\partial_\mu \lambda^I\,\partial^\mu\nabla^2
R\,,\qquad
\st_{16,17,18}=\cI_I^{5,6,7}\,\partial_\mu \lambda^I\,\nabla_\nu
  H_{2,3,4}^{\mu\nu}\,,
\end{gathered}}[]
and
\eqn{\begin{gathered}
\sz_{8}^\mu=\cG_I^1\,\partial^\mu \lambda^I\,E_4,\qquad
\sz_{9}^\mu=\cG_I^2\,\partial^\mu \lambda^I\,I,\qquad
\sz_{10}^\mu=\tfrac{1}{25}\cG_I^3\,\partial^\mu \lambda^I\,R^2,\\
\sz_{11}^\mu=\tfrac{1}{5}\cG_I^4\,\partial^\mu \lambda^I\,\nabla^2 R,\qquad
\sz_{12,\ldots,17}^\mu=\cH_I^{1,\ldots,6}\,\partial_\nu \lambda^I
  H_{1,\ldots,6}^{\mu\nu}\,,\\
\sz_{18}^\mu=\cF_I\,\nabla_\kappa\partial_\lambda \lambda^I\,\nabla^\mu
  G^{\kappa\lambda}\,,\qquad
\sz_{19}^\mu=\tfrac15\cE_I\,\nabla^2 \lambda^I\,\partial^\mu R\,.
\end{gathered}}[]
With more $\partial\lambda$'s we have
\eqn{\begin{gathered}
\st_{19}=\tfrac12\cG_{IJ}^1\,\partial_\mu \lambda^I\partial^\mu \lambda^J\,
E_4\,,\qquad
\st_{20}=\tfrac12\cG_{IJ}^2\,\partial_\mu \lambda^I\partial^\mu \lambda^J\,
I\,,\qquad
\st_{21}=\tfrac{1}{50}\cG_{IJ}^3\,\partial_\mu \lambda^I\partial^\mu
\lambda^J\,  R^2\,,\\
\st_{22}=\tfrac{1}{10}\cG_{IJ}^4\,\partial_\mu \lambda^I\partial^\mu
\lambda^J\,  \nabla^2 R\,,\qquad
\st_{23,\ldots,28}=\tfrac12\cH_{IJ}^{1,\ldots,6}\,\partial_\mu
  \lambda^I\partial_\nu \lambda^J\,H_{1,\ldots,6}^{\mu\nu}\,,\\
\st_{29}=\cF_{IJ}\,\partial_\kappa \lambda^I\nabla_\lambda\partial_\mu \lambda^J\,
  \nabla^\kappa G^{\lambda\mu}\,,\qquad
\st_{30}=\cF_{IJ}^\prime\,\partial_\kappa \lambda^I\nabla_\lambda\partial_\mu
  \lambda^J\,\nabla^\lambda G^{\kappa\mu}\,,
\end{gathered}}[]
and
\eqn{\begin{gathered}
\sz_{20}^\mu=\tfrac15\cE_{IJ}\,\partial^\mu \lambda^I\partial_\nu
  \lambda^J\,\partial^\nu R\,,\qquad
\sz_{21}^\mu=\cD_{IJ}\,\partial_\kappa \lambda^I\nabla_\lambda\partial_\nu
\lambda^J\,R^{\mu\lambda\kappa\nu}\,,\\
\sz_{22}^\mu=\cC_I\,\partial_\nu\nabla^2 \lambda^I\,G^{\mu\nu},\qquad
\sz_{23}^\mu=\cC_{IJ}\,\partial_\kappa \lambda^I\nabla_\nu
  \partial^\kappa \lambda^J\,G^{\mu\nu},\qquad
\sz_{24}^\mu=\cC_{IJ}^\prime\,\partial_\nu \lambda^I\nabla^2
\lambda^J\,G^{\mu\nu},\\
\sz_{25}^\mu=\tfrac15\cB_{IJ}\,\partial^\mu \lambda^I\nabla^2
\lambda^J\,R\,\qquad
\sz_{26}^\mu=\cA_{IJ}\,\partial_\nu\nabla^2 \lambda^I \nabla^\mu\partial^\nu
  \lambda^J,\qquad
\sz_{27}^\mu=\cA_{IJ}^{\prime}\,\partial^\mu \lambda^I(\nabla^2)^2
\lambda^J\,.
\end{gathered}}[]
Furthermore, we have
\eqn{\begin{gathered}
\st_{31}=\tfrac12\cF_{IJK}\,\partial_\kappa \lambda^I\partial_\lambda
  \lambda^J\partial_\mu \lambda^K\,\nabla^\kappa G^{\lambda\mu},\qquad
\st_{32}=\tfrac15\hat{\cE}_{IJ}\,\partial_\mu \lambda^I\nabla^2
\lambda^J\,\partial^\mu R\,,\\
\st_{33}=\tfrac{1}{10}\cE_{IJK}\,\partial_\mu \lambda^I\partial_\nu \lambda^J
  \partial^\nu \lambda^K\,\partial^\mu R\,,\qquad
\st_{34}=\cD_{IJK}\,\partial_\kappa \lambda^I \partial_\mu \lambda^J \nabla_\lambda
  \partial_\nu \lambda^K\,R^{\kappa\lambda\mu\nu},\\
\st_{35}=\tfrac14\cD_{IJKL}\,\partial_\kappa \lambda^I \partial_\lambda \lambda^J
  \partial_\mu \lambda^K\partial_\nu
  \lambda^L\,R^{\kappa\lambda\mu\nu},\qquad
\st_{36}=\hat{\cC}_{IJ}\,\nabla_\mu\partial_\nu \lambda^I\nabla^2
  \lambda^J\,G^{\mu\nu},\\
\st_{37}=\tfrac12\hat{\cC}_{IJ}^\prime\,\nabla_\kappa\partial_\mu \lambda^I
  \nabla^\kappa\partial_\nu \lambda^J\,G^{\mu\nu},\qquad
\st_{38}=\tfrac12\cC_{IJK}\,\partial_\mu \lambda^I\partial_\nu \lambda^J\nabla^2 \lambda^K\,
  G^{\mu\nu},\\
\st_{39}=\cC_{IJK}^\prime\,\partial_\mu \lambda^I\partial_\kappa \lambda^J\nabla^\kappa
  \partial_\nu \lambda^K\,G^{\mu\nu},\qquad
\st_{40}=\tfrac12\cC_{IJK}^{\prime\prime}\,\partial_\kappa \lambda^I
  \partial^\kappa \lambda^J \nabla_\mu\partial_\nu \lambda^K\, G^{\mu\nu},\\
\st_{41}=\tfrac14\cC_{IJKL}\,\partial_\mu \lambda^I \partial_\nu \lambda^J
  \partial_\kappa \lambda^K \partial^\kappa \lambda^L\,G^{\mu\nu},\qquad
\st_{42}=\tfrac15\cB_I\,(\nabla^2)^2 \lambda^I\,R\,,\\
\st_{43}=\tfrac{1}{10}\hat{\cB}_{IJ}\,\nabla^2 \lambda^I\nabla^2
\lambda^J\,R\,,\qquad
\st_{44}=\tfrac{1}{10}\hat{\cB}_{IJ}^\prime\,\nabla_\mu\partial_\nu \lambda^I
  \nabla^\mu\partial^\nu \lambda^J\,R\,,\\
\st_{45}=\tfrac{1}{10}\cB_{IJK}\,\partial_\mu \lambda^I\partial^\mu \lambda^J \nabla^2
  \lambda^K\,R\,,\qquad
\st_{46}=\tfrac{1}{10}\cB_{IJK}^\prime\,\partial_\mu \lambda^I\partial_\nu \lambda^J
  \nabla^\mu\partial^\nu \lambda^K\,R\,,\\
\st_{47}=\tfrac{1}{20}\cB_{IJKL}\,\partial_\mu \lambda^I\partial^\mu \lambda^J
  \partial_\nu \lambda^K\partial^\nu \lambda^L\,R\,,
\end{gathered}}[]
and
\eqn{\begin{gathered}
\sz_{28}^\mu=\cA_{IJK}\,\partial_\nu \lambda^I \nabla^\mu\partial^\nu
  \lambda^J\nabla^2 \lambda^K,\qquad
\sz_{29}^\mu=\cA_{IJK}^{\prime}\,\partial_\kappa \lambda^I
  \nabla^\mu\partial_\lambda \lambda^J\nabla^\kappa\partial^\lambda
  \lambda^K,\\
\sz_{30}^\mu=\tfrac12\cA_{IJKL}\,\partial_\nu \lambda^I \partial^\nu
  \lambda^J\partial^\mu \lambda^K \nabla^2 \lambda^L.
\end{gathered}}[]
Finally, we also have the terms
\eqn{\begin{gathered}
\st_{48}=\cA_I\,(\nabla^2)^3 \lambda^I,\qquad
\st_{49}=\hat{\cA}_{IJ}\,(\nabla^2)^2 \lambda^I\nabla^2 \lambda^J,\qquad
\st_{50}=\tfrac12\hat{\cA}_{IJ}^\prime\,\partial_\mu\nabla^2 \lambda^I\partial^\mu
  \nabla^2 \lambda^J,\\
\st_{51}=\tfrac12\hat{\cA}_{IJ}^{\prime\prime}\,\nabla_\kappa\nabla_\lambda
  \partial_\mu \lambda^I\nabla^\kappa\nabla^\lambda\partial^\mu \lambda^J,\qquad
\st_{52}=\tfrac18\hat{\cA}_{IJK}\,\nabla^2 \lambda^I \nabla^2 \lambda^J \nabla^2 \lambda^K,\\
\st_{53}=\tfrac12\hat{\cA}_{IJK}^{\prime}\,\nabla_\kappa\partial_\mu \lambda^I
  \nabla^\kappa\partial_\nu \lambda^J \nabla^\mu\partial^\nu \lambda^K,\qquad
\st_{54}=\hat{\cA}_{IJK}^{\prime\prime}\,\partial_\mu \lambda^I\nabla^2
  \lambda^J\partial^\mu \nabla^2 \lambda^K,\\
\st_{55}=\check{\cA}_{IJK}\,\partial_\mu \lambda^I\nabla^\mu\partial_\nu \lambda^J
  \partial^\nu\nabla^2 \lambda^K,\qquad
\st_{56}=\tfrac12\check{\cA}_{IJK}^\prime\,\partial_\mu \lambda^I\partial^\mu \lambda^J
  (\nabla^2)^2 \lambda^K,\\
\st_{57}=\tfrac12\check{\cA}_{IJK}^{\prime\prime}\,\partial_\mu \lambda^I
  \partial_\nu \lambda^J \nabla^\mu\partial^\nu\nabla^2 \lambda^K,\qquad
\st_{58}=\tfrac14\hat{\cA}_{IJKL}\,\partial_\mu \lambda^I \partial^\mu \lambda^J
  \nabla^2 \lambda^K\nabla^2 \lambda^L,\\
\st_{59}=\tfrac14\hat{\cA}_{IJKL}^\prime\,\partial_\kappa \lambda^I
  \partial^\kappa \lambda^J \nabla_\mu\partial_\nu \lambda^K \nabla^\mu\partial^\nu
  \lambda^L,\qquad
\st_{60}=\tfrac12\hat{\cA}_{IJKL}^{\prime\prime}\,\partial_\kappa \lambda^I
  \partial_\lambda \lambda^J \nabla^\kappa\partial_\mu \lambda^K
  \nabla^\lambda\partial^\mu \lambda^L,\\
\st_{61}=\tfrac12\check{\cA}_{IJKL}\,\partial_\mu \lambda^I
  \partial_\nu \lambda^J \nabla^\mu\partial^\nu \lambda^K\nabla^2 \lambda^L,\qquad
\st_{62}=\tfrac12\check{\cA}^\prime_{IJKL}\,\partial_\kappa \lambda^I
  \partial_\lambda \lambda^J \partial_\mu
  \lambda^K\nabla^\kappa\nabla^\lambda\partial^\mu \lambda^L,\\
\st_{63}=\tfrac14\cA_{IJKLM}\,\partial_\mu \lambda^I \partial^\mu \lambda^J
  \partial_\nu \lambda^K \partial^\nu \lambda^L \nabla^2 \lambda^M,\qquad
\st_{64}=\tfrac14\cA_{IJKLM}^\prime\,\partial_\kappa \lambda^I \partial^\kappa
  \lambda^J \partial_\lambda \lambda^K \partial_\mu \lambda^L \nabla^\lambda\partial^\mu
  \lambda^M,\\
\st_{65}=\tfrac18\cA_{IJKLMN}\,\partial_\kappa \lambda^I \partial^\kappa \lambda^J
  \partial_\lambda \lambda^K \partial^\lambda \lambda^L \partial_\mu \lambda^M \partial^\mu
  \lambda^N.
\end{gathered}}[]

\end{appendices}

\bibliography{6d_athm}

\end{document}